\documentclass[]{emulateapj}

\begin{document}

\newcommand{\cl}{{RDCS~1252.9-2927}}
\newcommand{\ms}{{MS~1358+62}}
\newcommand{\lynx}{{RDCS~0848+4453}}

\newcommand{\etal}{{\em et~al.\,}}
\title{The Fundamental Plane of Cluster Ellipticals at z = 1.25\altaffilmark{1,}\altaffilmark{2}}
\altaffiltext{1}{Based on observations with the NASA/ESA Hubble Space
Telescope, obtained at the Space Telescope Science Institute, which is
operated by the Association of Universities for Research in Astronomy,
Inc. under NASA contract No. NAS5-26555.}
\altaffiltext{2}{Based on observations obtained at the European
  Southern Observatory using the ESO Very Large Telescope on Cerro
  Paranal (ESO Programme 169.A-0458).}

\author{B. P. Holden\altaffilmark{3}}

\author{A. van der Wel\altaffilmark{4}}

\author{M. Franx\altaffilmark{4}}

\author{G. D. Illingworth\altaffilmark{3}}

\author{J. P. Blakeslee\altaffilmark{5}}

\author{P. van Dokkum\altaffilmark{6}}

\author{H. Ford\altaffilmark{5}}

\author{D. Magee\altaffilmark{3}}

\author{M. Postman\altaffilmark{5}}

\author{H.-W. Rix\altaffilmark{7}}

\author{P. Rosati\altaffilmark{8}}
\altaffiltext{3}{UCO/Lick Observatories, University of California,
  Santa Cruz, 95065; holden@ucolick.org; gdi@ucolick.org; magee@ucolick.org}
\altaffiltext{4}{Leiden Observatory, P.O.Box 9513, 2300 RA, Leiden,
  The Netherlands; vdwel@strw.leidenuniv.nl; franx@strw.leidenuniv.nl}
\altaffiltext{5}{Department of Physics \& Astronomy, Johns Hopkins University,
  Baltimore, MD 21218; jpb@pha.jhu.edu; ford@pha.jhu.edu; postman@stsci.edu}
\altaffiltext{6}{Yale University, New Haven, CT 06520-8101;
  dokkum@astro.yale.edu}
\altaffiltext{7}{Max-Planck-Institut f\"{u}r Astrnomie, K\"{o}nigstuhl
  17, D-69117, Heidelberg, Germany; rix@mpia.de}
\altaffiltext{8}{European Southern Observatory, Karl-Schwarzschild-Str. 2,
  D-85748, Garching, Germany; rosati@eso.org}

\shorttitle{The Cluster Fundamental Plane at z = 1.25}

\begin{abstract}

Using deep HST ACS imaging and VLT FORS2 spectra, we determined the
velocity dispersions, effective radii and surface brightnesses for
four early-type galaxies in the $z=1.237$ cluster \cl.  All four
galaxies are massive, $> 10^{11} M_{\sun}$.  These four galaxies,
combined with three from \lynx\ at $z=1.276$, establish the
Fundamental Plane of massive early-type cluster galaxies at
$\bar{z}=1.25$.  The offset of the Fundamental Plane shows that the
luminosity evolution in rest-frame $B$ is $\Delta \ln M/L_B = (-0.98
\pm 0.06) z$ for galaxies with $M > 10^{11.5} M_{\sun}$. To reproduce
the observed mass-to-light ratio ($M/L$) evolution, we determine the
characteristic age of the stars in these $M > 10^{11.5} M_{\sun}$
galaxies to be $3.0^{+0.3}_{-0.3}$ Gyrs, {\em i.e.}
$z_{*}=3.4^{+0.5}_{-0.4}$.  Including selection effects caused by
morphological bias (the ``progenitor bias''), we estimate an age of
$2.1^{+0.2}_{-0.2}$ Gyrs, or $z_{*} = 2.3^{+0.2}_{-0.2}$ for the
elliptical galaxy population.  Massive cluster early-type galaxies
appear to have a large fraction of stars that formed early in the
history of the universe.  However, there is a large scatter in the
derived $M/L$ values, which is confirmed by the spread in the
galaxies' colors.  Two lower mass galaxies in our $\bar{z}=1.25$
sample have much lower $M/L$ values, implying significant
star-formation close to the epoch of observation.  Thus, even in the
centers of massive clusters, there appears to have been significant
star formation in some massive, $M \simeq 10^{11} M_{\sun}$, galaxies
at $z\simeq 1.5$.

\end{abstract}

\section{Introduction}

The Fundamental Plane (FP) allows a direct measure of the mass and the
mass-to-light ratio, $M/L$, of early-type galaxies.  The FP combines
three variables, the effective radius ($r_e$), the average surface
brightness within the effective radius ($I_e$), and the velocity
dispersion ($\sigma$).  These data are combined into the relation \(
\sigma^{1.20} \propto r_e I_e^{0.83}\) for the rest-frame $B$ band
\citep{jfk96}.  With such quantities, we can measure $M/L \propto
\sigma^2/(r_e I_e) $ and how it depends on mass, $\propto r_e
\sigma^2$.  Massive galaxies out to $z \simeq 1$ appear to evolve as
$\Delta \ln M/L_B \simeq z $ for both clusters
\citep{kelson2000c,pvd_mf2001,wuyts2004} and in some field samples,
though there is a larger scatter for the latter
\citep{vandokkum2001b,gebhardt2003,vandokkum2003b,vandeven2003,vanderwel2004}.
This slow rate of evolution implies an early epoch of formation, $z_f
\simeq 3$, for the stars in early-type galaxies assuming passively
evolving simple stellar populations.  However, at $z \simeq 1.25$,
only $\sim50$\% of stellar mass we observe today have been formed
\citep[{\em e.g.}; ][]{madau98,steidel99,rudnick2003}.  This implies
that the majority of stars in cluster early-type galaxies formed long
before the average star in the universe.  Observations determining the
luminosity-weighted age of galaxies close to $z=1.25$ will test
this, and there are only three galaxies with FP measurements to date
at these redshifts \citep{vandokkum2003}.

We observed four luminous early-type galaxies in the $z=1.237$ rich,
massive and X-ray luminous cluster of galaxies \cl\ \citep{rosati2004}
using a combination of the Very Large Telescope (VLT) Focal
Reducer/low dispersion Spectrograph 2 (FORS2) and the Advanced Camera
for Surveys (ACS) on the Hubble Space Telescope (HST).  \cl\ is the
most massive cluster found to date at $z>1$, thus it contains a number
of luminous and, likely, massive galaxies.  We use the FP to constrain
the $M/L$ evolution and set the mass scale for four galaxies in \cl.
These results, combined with \citet{vandokkum2003}, measure the ages
of the stellar populations in early-type galaxies at $\bar{z}=1.25$.
We assume a $\Omega_{m} = 0.3$, $\Omega_{\Lambda} = 0.7$ and ${\rm H_o
= 70\ km\ s^{-1}\ Mpc^{-1}}$.  All observed magnitudes are in the AB
system.  However, for comparison with previous work, we convert observed
magnitudes into rest-frame Johnson $B$ using the Vega zeropoint.

\section{\cl\ Fundamental Plane Data}

\begin{figure}[htbp]
\begin{center}
\includegraphics[width=3.2in]{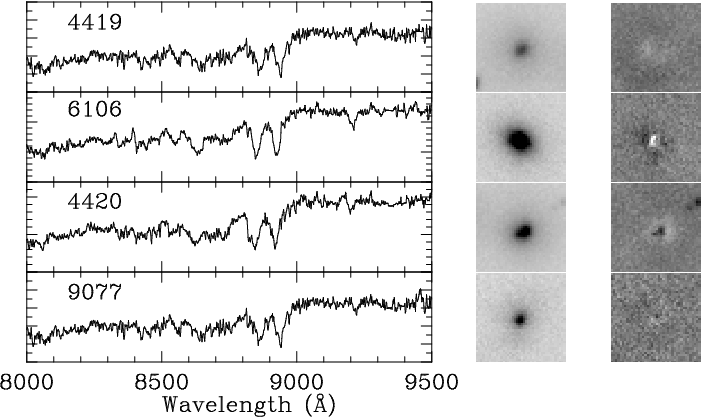}
\end{center}
\caption[f1.ps]{Mosaic showing, from left to right, the observed
  spectra near 4000\AA\ in the rest-frame of the cluster, images and
  images of the residuals.  Each image is 2\farcs 0 on a side with
  0\farcs 05 pixels from the $z_{850}$ ACS data, corresponding to
  roughly rest-frame Johnson $B$.  The fit results are listed in Table
  \ref{fits}.  The absolute value of flux in the residuals is
  $\le$10\% of the flux in the original data.  For \#4419 and \#4420,
  the residuals have an ``S'' shape which is interpreted as a sign of
  interaction \citep{blakeslee2003}.  The residuals for \#6106 show an
  over subtraction in the center, it is likely that the $r^{1/4}$ law
  is too steep; in this case the best fitting profile is $\propto
  r^{1/3}$. }
\label{mosaic}
\end{figure}

Four galaxies in \cl\ were selected from among the nine known cluster
members \citep{demarco2003PhDT,lidman2004} with $z_{850} < 22.5$ mag
that fit into a single multi-slit mask.  These are the first, second,
third and fifth brightest cluster members.  An image of each is shown
in Figure \ref{mosaic} along with the spectra, in descending order of
brightness.  Below we discuss the measurement of $r_e$ and $I_e$ from
the ACS data, the $\sigma$ from the FORS2 spectra and the conversion
to the rest-frame Johnson $B$.

\subsection{HST ACS Imaging}
\label{acs}

As described in \citet{blakeslee2003}, the Advanced Camera for Surveys
(ACS) imaged \cl\ with four overlapping pointings.  Each pointing has
three orbits in the F775W filter, or $i_{775}$, and five orbits in the
F850LP filter, or $z_{850}$.  A de Vaucouleurs, or $r^{1/4}$ model,
was fit to each of the individual $z_{850}$ images using the method of
\citet{vandokkum96}.  Each galaxy in each image had a unique point
spread function generated using the {\tt TinyTIM v6.2} package
\citep{krist95}.  The two galaxies at the middle of the cluster,
referred to as \# 4419 and \# 4420 in Table \ref{fits}, were fit
simultaneously. Table \ref{fits} contains the average of the best
fitting parameters for each image with the galaxies listed in order of
decreasing $z_{850}$ flux.  We plot in Figure \ref{mosaic} a mean
image, corrected for the ACS distortion, of all the $z_{850}$ images
for each galaxy along with the average residuals from the fits.


The product $r_e I_e^{0.83}$ is used to measure the evolution in $M/L$.
Because of the strong anti-correlation between the error for $\mu_e$
and the error for $r_{e}$ \citep{jfk93}, the uncertainties on this
product for all four galaxies is small at $\simeq$5\%.

\subsection{VLT FORS2 Spectra}
\label{spectra}


The four galaxies in Table \ref{fits} were observed using FORS2, on
the VLT, through slit masks with the 600z grism in conjunction with
the OG590 order separation filter.  The observations were done in
service mode with a series of exposures, dithered over four positions,
for a total integration time of 24 hours. The resulting
signal-to-noise (S/N) ratios at 4100 \AA\ rest-frame are listed in
Table \ref{fits}.  Details concerning the data reduction are described
in \citet{vanderwel2004b}.

The high spectral resolution, 80 ${\rm km\ s^{-1}}$ per pixel,
resulted in accurate internal velocity dispersions for the four
cluster members (see Table \ref{fits}).  The spectra were fit, by the
method of \citet{vandokkum96}, with stellar spectra \citep{valdes2004}
with a wide range in spectral type and metallicity. For more details
concerning the usage of the templates and the derivation of velocity
dispersions, see \citet{vanderwel2004b}.

The velocity dispersions were aperture corrected to a 1.7 kpc
circular aperture at the distance of Coma as described in
\citet{jfk96}.  The listed errors include a statistical error derived
from the $\chi^2$ value of the fit and a systematic error estimated to
be at most 5\% for the spectra with the lowest S/N ratio.

\subsection{Rest-frame Magnitudes}
\label{rest}

In order to compare with other FP results in the literature, the
observed $z_{850}$ magnitudes must be converted into $B_{rest}$, the
equivalent of observing the galaxies with a rest-frame Johnson B
filter \citep{bessell90} in the Vega system.  The $z_{850}$ filter is
centered at 4058\AA\ in the rest-frame of the galaxies in \cl. This
filter is close to the central wavelength of Johnson B at 4350\AA, but
even the modest wavelength difference means that the conversion
between $z_{850}$ at $z=1.237$ and the Johnson B will depend on the
color of the galaxy.  To compute this conversion, we redshifted the
Sbc and E templates from \citet{cww80} to $z=1.237$ and calculated the
$B_{rest}$ magnitude as a function the observed $z_{850}$ and
$i_{775}-J$ color, yielding $B_{rest} = z_{850} - 0.45 (i_{775} - J) +
1.68$.  This approach is slightly different than that used in
\citet{kelson2000a} or \citet{vandokkum2003}, but yields results that
differ in the mean by $\le 0.02$ magnitudes, with an error of only
$\le 5$\% \citep[see][]{holden2004}.  The $J$, along with $K_s$,
photometry comes from the VLT ISAAC and NTT SOFI observations
discussed in \citet{lidman2004}.  We will also use this data to
examine the rest-frame optical $B-I$ colors below.  All colors were
measured within an aperture of two effective radii.  The ACS imaging
was smoothed to match the seeing in the ISAAC data to measure this
color.

The statistical errors on the FP are dominated by the error on the
velocity dispersions, which are around 10\% including an estimate of
the systematic error. Because this error dominates the error budget,
we will take the error on $\sigma$ to be the FP error for the rest of
the paper.

\section{The Fundamental Plane at $\bar{z}=1.25$}

\begin{figure}[tbp]
\begin{center}
\includegraphics[width=3.2in]{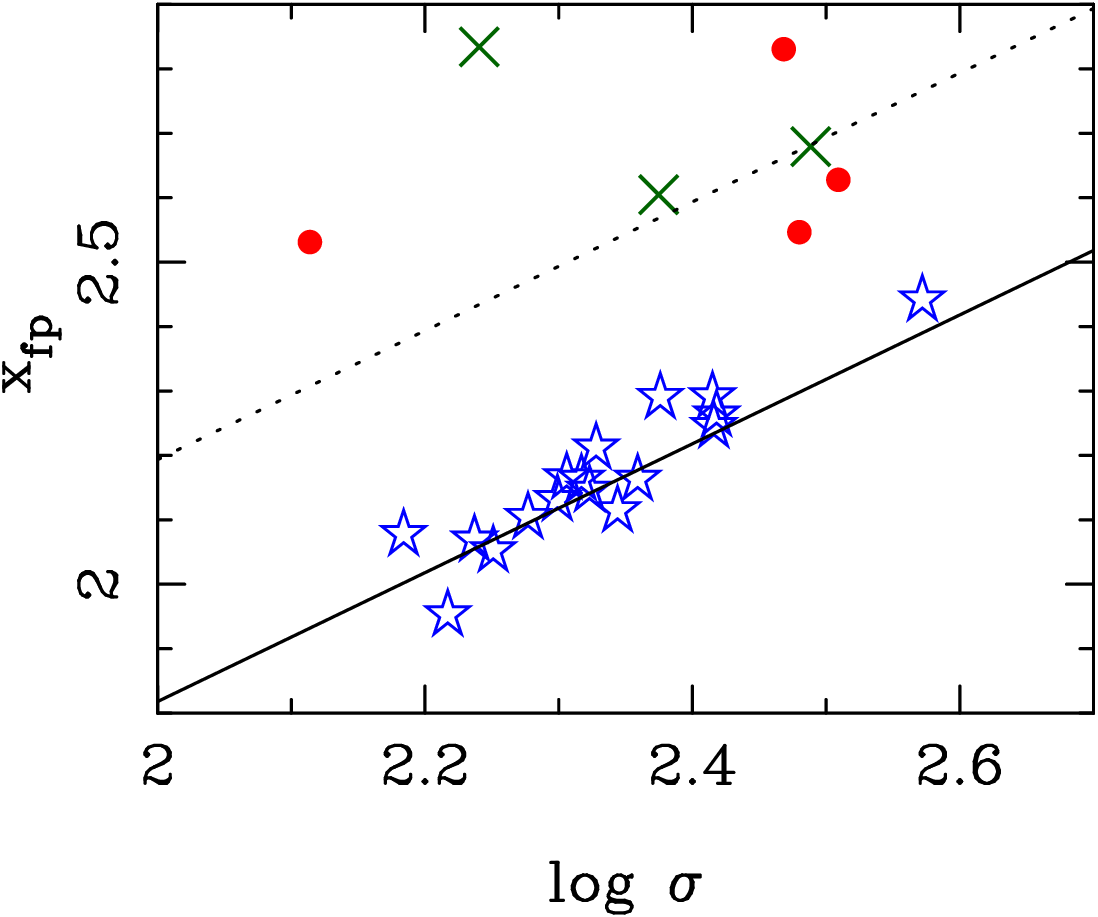}
\end{center}
\caption[f2.eps]{A projection of the FP for our sample and Coma.  The
  results for \cl\ (z=1.237) are plotted as solid red dots, the green
  crosses are from \lynx\ \citep[$z=1.276$]{vandokkum2003}, and the
  blue open stars are from Coma \citep[$z=0.023$]{jfk96}.  The y-axis
  is $x_{fp} = 0.83 \log r_e + 0.69 \log I_e$, one of the projections
  of the FP.  The solid line is the FP for Coma while the dotted line
  has the same slope but is shifted $\Delta M/L_B = -0.98$ for the
  five massive galaxies at $\bar{z}=1.25$. }
\label{fpevol}
\end{figure}

\begin{figure}[tbp]
\begin{center}
\includegraphics[width=3.2in]{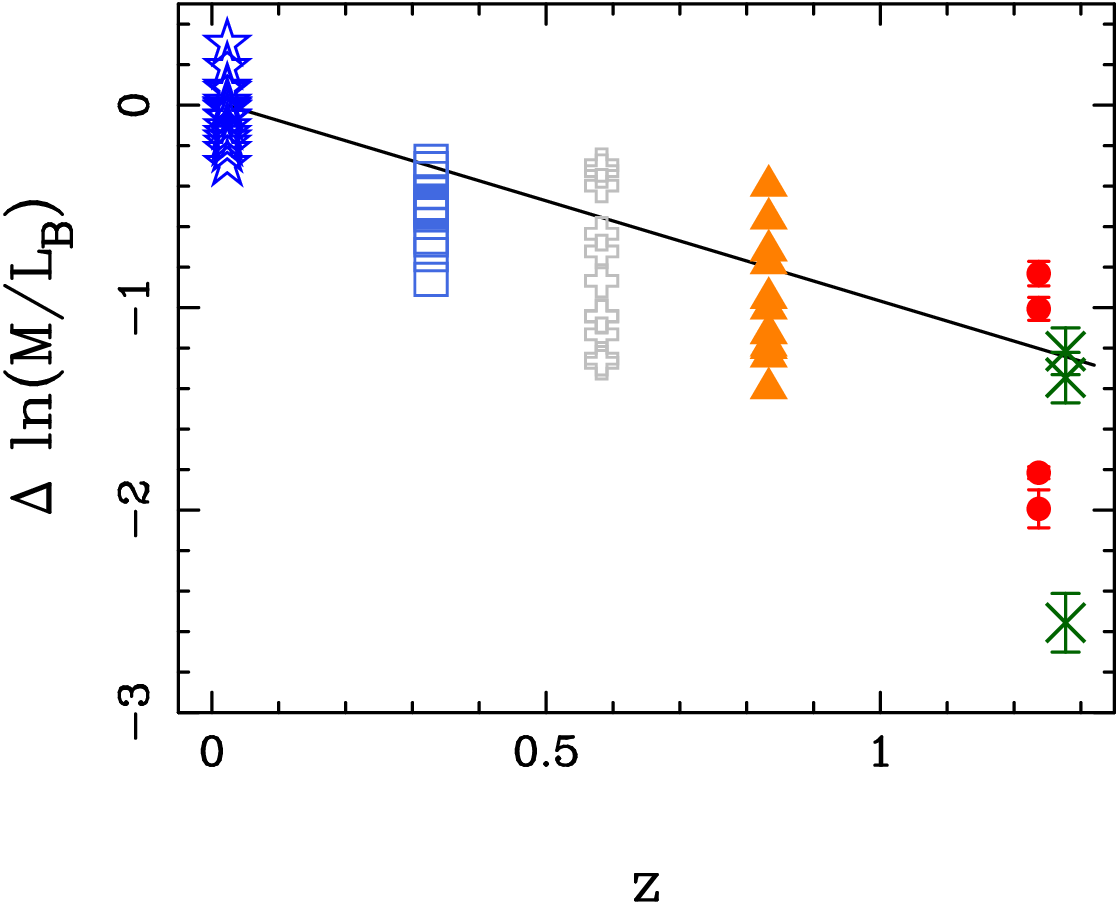}
\end{center}
\caption[f3.eps]{Change in $M/L_B$ for early-types as a function of
  redshift.  We use the same symbols as Figure \ref{fpevol} with the
  addition of light blue open squares for \ms\
  \citep[$z=0.328$]{kelson2000c}, grey open ``$+$'' are from
  MS~2053-04 ($z=0.583$) and the filled orange triangles are from
  MS~1054-03 ($z=0.832$), both from \citet{wuyts2004}.  The resulting
  evolution, with respect to the Coma FP, is shown as a line with the
  form $\Delta \ln M/L_B \propto (-0.98 \pm 0.06)\, z$.  }
\label{mlevol}
\end{figure}

The most straightforward way to measure the evolution in $M/L_B$ is to
compute the offsets for the seven galaxies at $\bar{z}=1.25$ from the
FP of the Coma cluster \citep{jfk96}. In Figure \ref{fpevol}, we plot
our data along, with the FP for Coma, using the rest-frame $B$.  We
show the average offset for the five most massive galaxies out of the
total of seven in our sample.  We will discuss below why we remove
those two lower mass galaxies.  Measuring the offset from the FP
ensures that we measuring $\Delta M/L_B$ for galaxies at the same part
of the FP or, roughly the same mass.  The offset $\Delta M/L$ and rate
of $M/L_B$ evolution is readily apparent in Figure \ref{mlevol}. We
find $\Delta \bar{M/L_B} = -1.23 \pm 0.08$ for five $\bar{z}=1.25$
early-type galaxies, three from \cl\ and two from \lynx\ with masses
$M > 10^{11.5} M_{\sun}$. This corresponds to an evolution in $\Delta
\ln M/L_B \propto (-0.98 \pm 0.06)\, z$, a small deviation from the
$\Delta \ln M/L_B \propto (-1.06 \pm 0.09)\, z$ of
\citet{vandokkum2003} and the $\Delta \ln M/L_B \propto -1.08 \, z$ of
\citet{wuyts2004}.

There is a large scatter seen in Figure \ref{mlevol} in $\Delta \ln
M/L_B$, $\sigma (\ln M/L_B) = 0.32$ for the seven $\bar{z}=1.25$,
early-type galaxies.  This scatter is twice the size of the scatter in
Coma or \ms, regardless of whether the scatter is computed for all
galaxies, or only the seven most luminous galaxies in either \ms, or
the Coma cluster sample.  A large part of this scatter comes from the
two lower mass galaxies in the sample.  The five galaxies with $M >
10^{11.5} M_{\sun}$ show $\sigma (\ln M/L_B) = 0.22$, which is not
statistically different from the Coma or \ms\ value.  There is an
obvious selection effect towards low $M/L$ galaxies in a
luminosity-selected sample. This may both increase the scatter and
bias the mean change in the $M/L$ ratio, hence we remove the two low
mass galaxies from our sample.

\begin{figure}[tp]
\begin{center}
\includegraphics[width=3.2in]{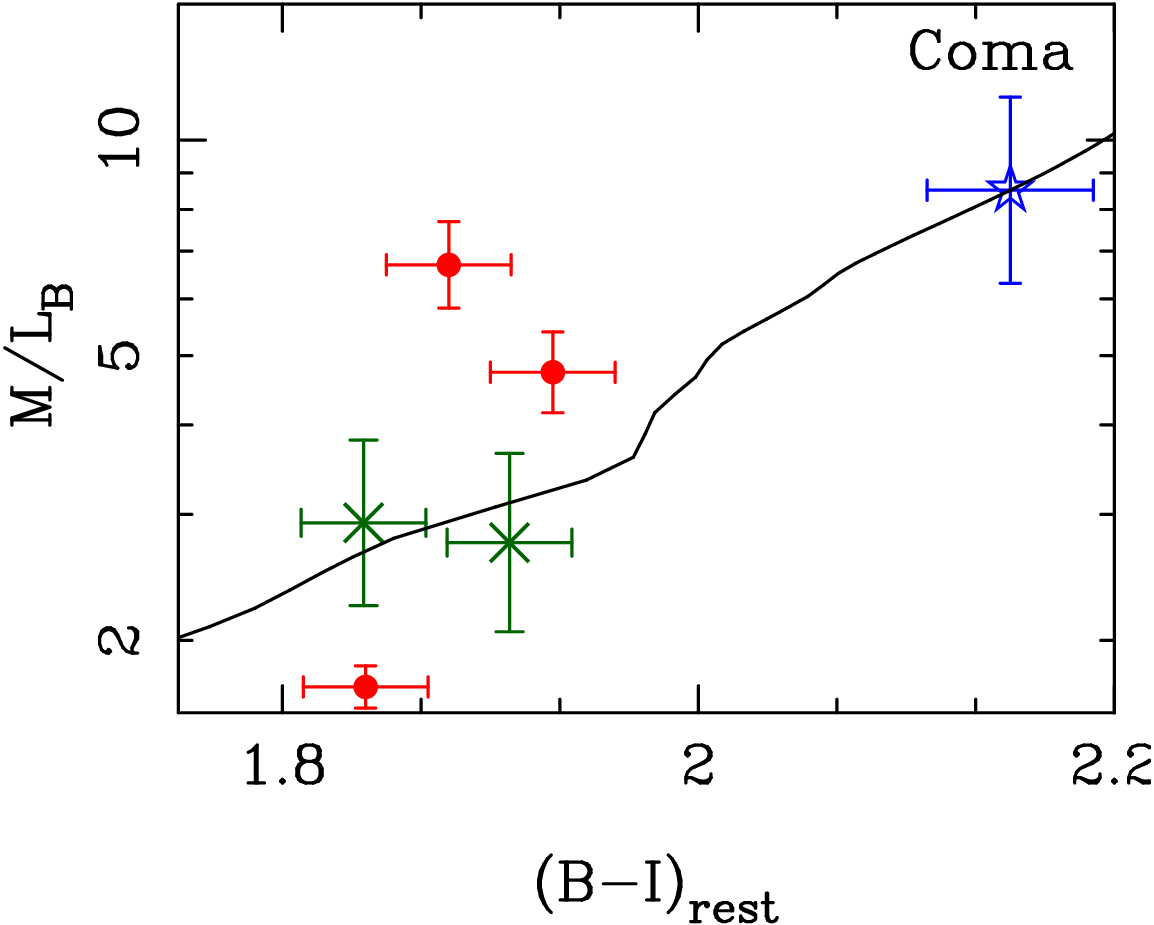}
\end{center}
\caption[f4.eps]{$M/L_B$ as a function of the rest-frame Vega $B-I$
  color for galaxies with $M > 10^{11.5} M_{\sun}$.  The average Coma
  values are represented as a blue star, while the three galaxies from
  \cl\ are solid circles and the two galaxyes from \lynx\ are green
  crosses.  The line shows the trajectory of a solar metallicity model
  with an exponentially declining, $\tau = 200$ Myrs, star-formation
  rate \cite{bc03} model normalized to the average Coma cluster
  observations.  The highest $M/L_B$ is is \#4419, the brightest
  cluster galaxy.  The colors of \#4419 are significantly bluer than
  predicted for its observed $M/L_B$, ruling out the offset being a
  result of dust or metallicity effects.  The lowest mass galaxy in
  \lynx\ is excluded (see text). }

\label{color_evol}
\end{figure}

The $M/L_B$ values for the $M > 10^{11.5} M_{\sun}$ galaxies at
$\bar{z}=1.25$ correlate with the rest-frame $B-I$ colors, as seen in
Figure \ref{color_evol}.  Both the colors and the $M/L_B$ track a
rapidly declining star-formation rate model from \citet{bc03}.  As
this relatively simple stellar population reproduces most of the
observations, the observed scatter in $M/L_B$ is then likely the
result of a spread in the luminosity-weighted ages.  

\begin{figure}[tbp]
\begin{center}
\includegraphics[width=3.2in]{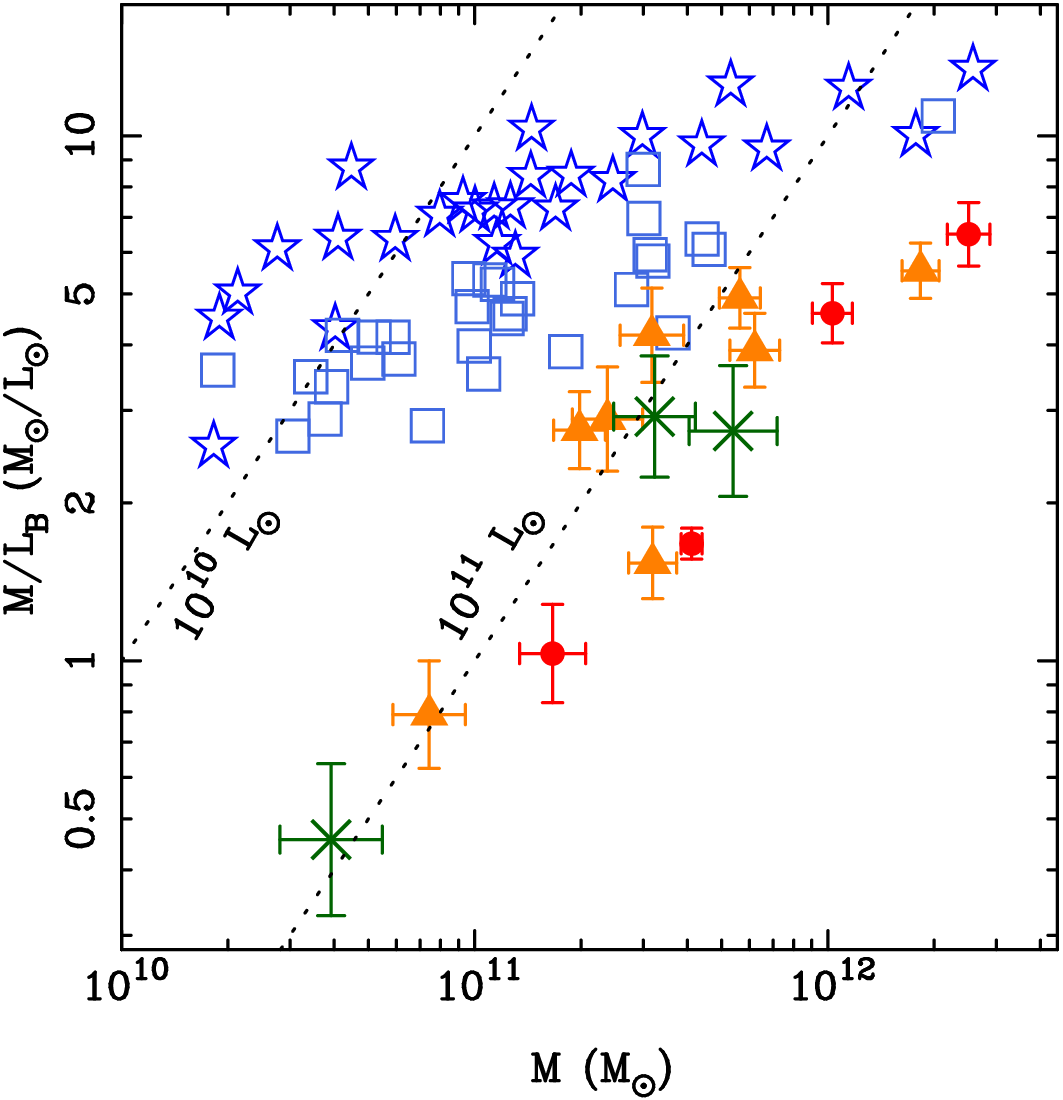}
\end{center}
\caption[f4.eps]{$M/L_B$ as a function of mass, using the same symbols
  as Figure \ref{mlevol} but not including the results for MS~2053-04.
  The lower redshift trend of higher $M/L_B$ at higher masses appears
  to be preserved at $\bar{z} = 1.25$, even when ignoring the lowest
  mass galaxy in \lynx.  The dotted lines represent $L_B = 10^{11}\,
  L_{\sun}$ and $10^{10}\, L_{\sun}$.  Our $z_{850} = 22.5$ selection
  limit corresponds to $L_B \simeq 10^{10.9} L_{\sun}$ for a galaxy
  with the colors of a Coma elliptical.}
\label{ml}
\end{figure}

At lower redshifts, the $M/L$ is a function of the total mass
\citep{jfk96}.  Such a trend is observed at $\bar{z}=1.25$ (see Figure
\ref{ml}) where lower mass galaxies have lower $M/L_B$ and, therefore,
bluer colors (see Figure \ref{color_evol}.)  The slope of the
mass-$M/L$ relation appears to steepen at higher redshifts.  This is
expected for a population where the spread in the $M/L$ comes from the
spread in age.  As a stellar population becomes younger, the $M/L$
changes more quickly, so the spread in $M/L$ will grow as observations
probe closer to the epoch of formation.  However, this trend will be
exaggerated by the selection of galaxies at a fixed luminosity as the
sample selection will prefer lower $M/L$ galaxies.  Lower mass
galaxies with larger values for $M/L_B$ will not appear in this sample
because of our magnitude limit, effectively $ L_B \simeq 10^{11}
L_{\sun} $.

\section{Implications for the Evolution of Early-type Cluster Galaxies}

The rate of the $M/L$ evolution can be used to constrain the
luminosity-weighted age for early-type galaxies.  The most
straight-forward estimate is to assume all galaxies formed at one
epoch and find the age of the galaxies that will produce the observed
change in $M/L$ from $z=0.023$ to $\bar{z}=1.25$ using population
synthesis models.  For our sample of five $M > 10^{11.5} M_{\sun}$
galaxies, the mean age is $\tau_{*} = 3.0^{+0.3}_{-0.3}$ Gyrs before
the time of the observations, or a formation redshift of $z_{*} =
3.4^{+0.5}_{-0.4}$ using the same models as \citet{vandokkum2003},
namely \citet{worthey1994}.  This agrees with the results from the high
mass sample of \citet{wuyts2004}, who found $z_{*} =
2.95^{+0.81}_{-0.46}$ for galaxies regardless of morphology.

When computing the age for the early-type galaxy population, there is
an overestimate of the age of a population caused by young galaxies
not being counted as part of the early-type population, even if those
young galaxies will evolve into early-types after the epoch of
observation.  This ``progenitor bias'' depresses the rate of observed
evolution in $M/L$ by up to 20\% \citep{pvd_mf2001}.  Using this same
assumption, namely that the true evolution is $\Delta \ln M/L_B
\propto (-1.18 \pm 0.06)\, z$, we instead find $\tau_{*} =
2.1^{+0.2}_{-0.2}$ Gyrs before $\bar{z}=1.25$, or a formation redshift
of $z_{*} = 2.3^{+0.2}_{-0.2}$.  \citet{blakeslee2003} and
\citet{lidman2004} both find a mean age of $\ge 2.6$ Gyrs using the
colors of the galaxies in \cl.  \citet{blakeslee2003} removed the 
``progenitor bias'' with simulations, whereas \citet{lidman2004} uses
all galaxies, regardless of morphology to similar effect.


The above results imply that the stars that formed these massive
galaxies were created at redshifts of $z_{*} \simeq 2-3$, at which
time less than $1/3$ of today's observed stellar mass was formed
\citep[{\em e.g.};][]{bell2005}.  However, there is a large spread in
the $M/L_B$ for all of the early-type galaxies at $z \simeq 1.25$,
larger than at lower redshifts.  Using colors confirms the spread,
which can be interpreted as an underlying spread in the age of the
populations.  Thus, though $z_{*} \simeq 2.5$ for the most massive
galaxies, some early-type galaxies show much lower $M/L$ values and,
corresponding younger ages.  In fact, the lowest $M/L$ galaxy in
Figure \ref{mlevol} was tentatively classified by
\citet{vandokkum2003} as having a recent star-burst based on the
spectrum.  Such younger appearing galaxies have lower masses than the
high $M/L$ galaxies in our sample, but are still massive galaxies with
$\log M/M_{\sun} \simeq 11$.  The implication of all these results is
that a significant fraction of the stars in the most massive galaxies
appear to have formed very early in the history of the universe,
before the majority of stars present today.  The massive cluster
galaxies appear to follow the same low-redshift trend of the higher
mass systems having higher $M/L_B$.  However, the larger spread at
$\bar{z} = 1.25$ in the $M/L_B$ indicates that we have identified some
massive, $M \simeq 10^{11} M_{\sun} $, galaxies whose last burst of
star formation occurred in the relatively recent past, $z\simeq 1.5$.

ACS was developed under NASA contract NAS5-32865, and this research
was supported by NASA grant NAG5-7697.  BH would like to thank Daniel
Kelson for useful discussions on the Fundamental Plane.  The authors
would also like to thank the referee for many useful suggestions.


\begin{thebibliography}{29}
\expandafter\ifx\csname natexlab\endcsname\relax\def\natexlab#1{#1}\fi

\bibitem[{{Bell}(2005)}]{bell2005}
{Bell}, E.~F. 2005, in Planets to Cosmology: Essential Science in Hubble's
  Final Years, ed. M.~{Livio} (Cambridge: CUP), in press, astro--ph/0408023

\bibitem[{{Bessell}(1990)}]{bessell90}
{Bessell}, M.~S. 1990, \pasp, 102, 1181

\bibitem[{{Blakeslee} {et~al.}(2003){Blakeslee}, {Franx}, {Postman}, {Rosati},
  {Holden}, {Illingworth}, {Ford}, {Cross}, {Gronwall}, {Ben{\'{\i}}tez},
  {Bouwens}, {Broadhurst}, {Clampin}, {Demarco}, {Golimowski}, {Hartig},
  {Infante}, {Martel}, {Miley}, {Menanteau}, {Meurer}, {Sirianni}, \&
  {White}}]{blakeslee2003}
{Blakeslee}, J.~P., {Franx}, M., {Postman}, M., {Rosati}, P., {Holden}, B.~P.,
  {Illingworth}, G.~D., {Ford}, H.~C., {Cross}, N.~J.~G., {Gronwall}, C.,
  {Ben{\'{\i}}tez}, N., {Bouwens}, R.~J., {Broadhurst}, T.~J., {Clampin}, M.,
  {Demarco}, R., {Golimowski}, D.~A., {Hartig}, G.~F., {Infante}, L., {Martel},
  A.~R., {Miley}, G.~K., {Menanteau}, F., {Meurer}, G.~R., {Sirianni}, M., \&
  {White}, R.~L. 2003, \apjl, 596, L143

\bibitem[{{Bruzual} \& {Charlot}(2003)}]{bc03}
{Bruzual}, G. \& {Charlot}, S. 2003, \mnras, 344, 1000

\bibitem[{{Coleman} {et~al.}(1980){Coleman}, {Wu}, \& {Weedman}}]{cww80}
{Coleman}, G.~D., {Wu}, C.-C., \& {Weedman}, D.~W. 1980, \apjs, 43, 393

\bibitem[{{Demarco}(2003)}]{demarco2003PhDT}
{Demarco}, R. 2003, Ph.D.~Thesis

\bibitem[{{Gebhardt} {et~al.}(2003){Gebhardt}, {Faber}, {Koo}, {Im}, {Simard},
  {Illingworth}, {Phillips}, {Sarajedini}, {Vogt}, {Weiner}, \&
  {Willmer}}]{gebhardt2003}
{Gebhardt}, K., {Faber}, S.~M., {Koo}, D.~C., {Im}, M., {Simard}, L.,
  {Illingworth}, G.~D., {Phillips}, A.~C., {Sarajedini}, V.~L., {Vogt}, N.~P.,
  {Weiner}, B., \& {Willmer}, C.~N.~A. 2003, \apj, 597, 239

\bibitem[{{Holden} {et~al.}(2004){Holden}, {Stanford}, {Eisenhardt}, \&
  {Dickinson}}]{holden2004}
{Holden}, B.~P., {Stanford}, S.~A., {Eisenhardt}, P.~R., \& {Dickinson}, M.
  2004, \aj, 127, 2484

\bibitem[{{Jorgensen} {et~al.}(1993){Jorgensen}, {Franx}, \&
  {Kjaergaard}}]{jfk93}
{Jorgensen}, I., {Franx}, M., \& {Kjaergaard}, P. 1993, \apj, 411, 34

\bibitem[{{Jorgensen} {et~al.}(1996){Jorgensen}, {Franx}, \&
  {Kjaergaard}}]{jfk96}
---. 1996, \mnras, 280, 167

\bibitem[{{Kelson} {et~al.}(2000{\natexlab{a}}){Kelson}, {Illingworth}, {van
  Dokkum}, \& {Franx}}]{kelson2000a}
{Kelson}, D.~D., {Illingworth}, G.~D., {van Dokkum}, P.~G., \& {Franx}, M.
  2000{\natexlab{a}}, \apj, 531, 137

\bibitem[{{Kelson} {et~al.}(2000{\natexlab{b}}){Kelson}, {Illingworth}, {van
  Dokkum}, \& {Franx}}]{kelson2000c}
---. 2000{\natexlab{b}}, \apj, 531, 184

\bibitem[{{Krist}(1995)}]{krist95}
{Krist}, J. 1995, in ASP Conf. Ser. 77: Astronomical Data Analysis Software and
  Systems IV, 349

\bibitem[{{Lidman} {et~al.}(2004){Lidman}, {Rosati}, {Demarco}, {Nonino},
  {Mainieri}, {Stanford}, \& {Toft}}]{lidman2004}
{Lidman}, C., {Rosati}, P., {Demarco}, R., {Nonino}, M., {Mainieri}, V.,
  {Stanford}, S.~A., \& {Toft}, S. 2004, \aap, 416, 829

\bibitem[{{Madau} {et~al.}(1998){Madau}, {Pozzetti}, \& {Dickinson}}]{madau98}
{Madau}, P., {Pozzetti}, L., \& {Dickinson}, M. 1998, \apj, 498, 106

\bibitem[{{Rosati} {et~al.}(2004){Rosati}, {Tozzi}, {Ettori}, {Mainieri},
  {Demarco}, {Stanford}, {Lidman}, {Nonino}, {Borgani}, {Della Ceca},
  {Eisenhardt}, {Holden}, \& {Norman}}]{rosati2004}
{Rosati}, P., {Tozzi}, P., {Ettori}, S., {Mainieri}, V., {Demarco}, R.,
  {Stanford}, S.~A., {Lidman}, C., {Nonino}, M., {Borgani}, S., {Della Ceca},
  R., {Eisenhardt}, P., {Holden}, B.~P., \& {Norman}, C. 2004, \aj, 127, 230

\bibitem[{{Rudnick} {et~al.}(2003){Rudnick}, {Rix}, {Franx}, {Labb{\' e}},
  {Blanton}, {Daddi}, {F{\" o}rster Schreiber}, {Moorwood}, {R{\" o}ttgering},
  {Trujillo}, {van der Wel}, {van der Werf}, {van Dokkum}, \& {van
  Starkenburg}}]{rudnick2003}
{Rudnick}, G., {Rix}, H., {Franx}, M., {Labb{\' e}}, I., {Blanton}, M.,
  {Daddi}, E., {F{\" o}rster Schreiber}, N.~M., {Moorwood}, A., {R{\"
  o}ttgering}, H., {Trujillo}, I., {van de Wel}, A., {van der Werf}, P., {van
  Dokkum}, P.~G., \& {van Starkenburg}, L. 2003, \apj, 599, 847

\bibitem[{{Steidel} {et~al.}(1999){Steidel}, {Adelberger}, {Giavalisco},
  {Dickinson}, \& {Pettini}}]{steidel99}
{Steidel}, C.~C., {Adelberger}, K.~L., {Giavalisco}, M., {Dickinson}, M., \&
  {Pettini}, M. 1999, \apj, 519, 1

\bibitem[{{Valdes} {et~al.}(2004){Valdes}, {Gupta}, {Rose}, {Singh}, \&
  {Bell}}]{valdes2004}
{Valdes}, F., {Gupta}, R., {Rose}, J.~A., {Singh}, H.~P., \& {Bell}, D.~J.
  2004, \apjs, 152, 251

\bibitem[{{van de Ven} {et~al.}(2003){van de Ven}, {van Dokkum}, \&
  {Franx}}]{vandeven2003}
{van de Ven}, G., {van Dokkum}, P.~G., \& {Franx}, M. 2003, \mnras, 344, 924

\bibitem[{{van der Wel} {et~al.}(2004{\natexlab{a}}){van der Wel}, {Franx},
  {van Dokkum}, \& {Rix}}]{vanderwel2004b}
{van der Wel}, A., {Franx}, M., {van Dokkum}, P.~G., \& {Rix}, H.-W.
  2004{\natexlab{a}}, \apj, submitted

\bibitem[{{van der Wel} {et~al.}(2004{\natexlab{b}}){van der Wel}, {Franx},
  {van Dokkum}, \& {Rix}}]{vanderwel2004}
---. 2004{\natexlab{b}}, \apjl, 601, L5

\bibitem[{{van Dokkum} \& {Ellis}(2003)}]{vandokkum2003b}
{van Dokkum}, P.~G. \& {Ellis}, R.~S. 2003, \apjl, 592, L53

\bibitem[{{van Dokkum} \& {Franx}(1996)}]{vandokkum96}
{van Dokkum}, P.~G. \& {Franx}, M. 1996, \mnras, 281, 985

\bibitem[{{van Dokkum} \& {Franx}(2001)}]{pvd_mf2001}
---. 2001, \apj, 553, 90

\bibitem[{{van Dokkum} {et~al.}(2001){van Dokkum}, {Franx}, {Kelson}, \&
  {Illingworth}}]{vandokkum2001b}
{van Dokkum}, P.~G., {Franx}, M., {Kelson}, D.~D., \& {Illingworth}, G.~D.
  2001, \apjl, 553, L39

\bibitem[{{van Dokkum} \& {Stanford}(2003)}]{vandokkum2003}
{van Dokkum}, P.~G. \& {Stanford}, S.~A. 2003, \apj, 585, 78

\bibitem[{{Worthey}(1994)}]{worthey1994}
{Worthey}, G. 1994, \apjs, 95, 107

\bibitem[{{Wuyts} {et~al.}(2004){Wuyts}, {van Dokkum}, {Kelson}, {Franx}, \&
  {Illingworth}}]{wuyts2004}
{Wuyts}, S., {van Dokkum}, P.~G., {Kelson}, D.~D., {Franx}, M., \&
  {Illingworth}, G.~D. 2004, \apj, 605, 677

\end{thebibliography}

\begin{deluxetable*}{lrccccccc}
\tablecolumns{6}
\tablecaption{de Vaucouleurs Model Parameters and Velocity
  Dispersion\label{fits}} 
\tablehead{
\colhead{Galaxy} & \colhead{$r_{e,z}$} &
\colhead{$\mu_e$\tablenotemark{a}} & \colhead{S/N\tablenotemark{b}} &
\colhead{$\sigma$} & $(i-J)_{2r_e}$\tablenotemark{a}& $\log$ Mass &
$\log M/L_B$ \\ 
\colhead{} & \colhead{(\arcsec)} & \colhead{(mag
 per \sq\arcsec)} & \colhead{\AA$^{-1}$} &
 \colhead{(${\rm km\ s^{-1}}$)} & \colhead{mag}
& $\log M_{\sun}$ &  
 \\
}
\startdata
4419 & 2.806 & 24.899 & 24 & 302 $\pm$ 24 & 2.09 $\pm$ 0.02
& 12.40 $\pm$ 0.08 & 0.81 $\pm$ 0.06 \\
6106 & 0.487  & 21.573 & 57 & 294 $\pm$ 10 & 2.07 $\pm$ 0.02
& 11.61 $\pm$ 0.08 & 0.22 $\pm$ 0.03 \\
4420 & 1.016  & 23.279 & 29 & 323 $\pm$ 21 & 2.11 $\pm$ 0.02
& 12.01 $\pm$ 0.06 & 0.66 $\pm$ 0.06 \\
9077 & 1.008  & 23.529 & 24 & 130 $\pm$ 14 & 1.90 $\pm$ 0.02
& 11.22 $\pm$ 0.08 & 0.01 $\pm$ 0.09 \\
\enddata
\tablenotetext{a}{All magnitudes are AB.}
\tablenotetext{b}{Spectra have a resolution of 3.7 \AA (FWHM).}
\end{deluxetable*}

\end{document}